\documentclass[aps,prd,twocolumn,superscriptaddress,footinbib]{revtex4-1}
\usepackage[utf8]{inputenc}
\usepackage[colorlinks=true,linkcolor=blue,citecolor=blue,urlcolor=blue]{hyperref}
\usepackage{bm,bbm,amssymb,amsmath}
\usepackage{graphicx}
\usepackage[dvipsnames]{xcolor}
\usepackage{comment}
\usepackage{url}
\usepackage{footnote}
\newcommand{\UofT}{Department of Physics, University of Toronto, Toronto, Ontario M5S 1A7, Canada}
\newcommand{\CITA}{Canadian Institute for Theoretical Astrophysics, University of Toronto, Toronto, Ontario M5S 3H8, Canada}
\newcommand{\NITR}{Department of Physics \& Astronomy, National Institute of Technology, Rourkela 769008, India}
\newcommand{\PI}{Perimeter Institute for Theoretical Physics, Waterloo, Ontario N2L 2Y5, Canada}
\newcommand{\INFN}{INFN Sezione di Catania, Dipartimento di Fisica,Via S. Sofia 64, 95123 Catania, Italy}

\begin{document}
\title{Astrophysical constraints on neutron star $f$-modes \\ with a nonparametric equation of state representation}
\author{Sailesh Ranjan Mohanty}
\affiliation{\NITR}
\email{saileshranjanmohanty@gmail.com}
\author{Utkarsh Mali}
\affiliation{\CITA}
\affiliation{\UofT}
\email{utkarsh.mali@utoronto.ca}
\author{H.C. Das}
\affiliation{\INFN}
\author{Bharat Kumar}
\affiliation{\NITR}
\author{Philippe Landry}
\affiliation{\CITA}
\affiliation{\PI}
\begin{abstract}
We constrain the fundamental-mode ($f$-mode) oscillation frequencies of nonrotating neutron stars using a phenomenological Gaussian process model for the unknown dense-matter equation of state conditioned on a suite of gravitational-wave, radio and X-ray observations. We infer the quadrupolar $f$-mode frequency preferred by the astronomical data as a function of neutron star mass, with error estimates that quantify the impact of equation of state uncertainty, and compare it to the contact frequency for inspiralling neutron-star binaries, finding that resonance with the orbital frequency can be achieved for the coalescences with the most unequal mass ratio. For an optimally configured binary neutron star merger, we estimate the gravitational waveform's tidal phasing due to $f$-mode dynamical tides as $7^{+2}_{-3}$ rad at merger. We assess prospects for distinguishing $f$-mode dynamical tides with current and future-generation gravitational-wave observatories.
\end{abstract}
\maketitle
\section{Introduction} \label{sec:intro}

Dynamical tides are an important aspect of the tidal interaction of neutron-star binaries during gravitational-wave driven inspiral. The neutron star oscillation mode with the strongest tidal coupling is the fundamental mode, or $f$-mode, the longest-wavelength pressure-restored mode with no radial nodes~\cite{KokkotasSchmidt1999}. The frequency of the $f$-mode depends on the neutron star's mass and its internal structure, as encoded in its equation of state (EOS). During binary inspiral, $f$-mode oscillations can exchange energy with the orbit when the orbital frequency approaches or achieves resonance with the mode frequency~\cite{Lai1994,HoLai1999}. Fundamental modes are also an important oscillation mode in post-merger remnants of binary neutron star mergers~\cite{StergioulasBauswein2011,BausweinJanka2012}, where the (nonrotating, zero-temperature, barotropic) $f$-mode frequency of the inspiral is corrected by rotational, thermal and compositional effects~\cite{LioutasBauswein2021,RaithelPaschalidis2024,ChabanovRezzolla2023}. While $f$-mode oscillations have not yet been measured directly, neglecting their impact on the tidal phasing of the gravitational waves from neutron star mergers---even away from resonance---can bias the recovery of adiabatic tidal effects captured by the tidal deformability parameter $\Lambda$~\cite{PrattenSchmidt2022}. To mitigate this bias, $f$-mode dynamical tides have been incorporated into waveform models~\cite{HindererTaracchini2016,SteinhoffHinderer2016,SchmidtHinderer2019,LackeyPurrer2019,SteinhoffHinderer2021,GambaBernuzzi2023,AbacDietrich2024}. In some of these models~\cite{LackeyPurrer2019,AbacDietrich2024}, the uncertain---and presently unmeasurable~\cite{PrattenSchmidt2020}---$f$-mode frequency is fixed by an empirical relation linking it to $\Lambda$~\cite{ChanSham2014,ChirentideSouza2015,ZhaoLattimer2022}. 
This empirical relation is based on a quasi-universal correlation exhibited by candidate equations of state from nuclear theory.

An alternative approach to fixing the $f$-mode frequency is to directly marginalize, or average, its value over a distribution that captures the uncertainty in the EOS. To enable this treatment, here we use a large ensemble of observationally  conditioned phenomenological nonparametric EOS realizations~\cite{LegredChatziioannou2021} to compute a posterior distribution over the quadrupolar $f$-mode frequencies $f_2$ of nonrotating neutron stars. This data-driven, nonparametric approach contrasts with other studies~\cite{JaiswalChatterjee2021,KeshariPradhanChatterjee2022} that work within a particular microscopic framework for the EOS, such as relativistic mean-field theory, or with a specific EOS parameterization~\cite{GuhaRoyMalik2024}. Equipped with a best estimate of $f_2$ for a neutron star of a given mass $m$, including errors that encode EOS uncertainty, we address several questions about $f$-mode oscillations in inspiralling neutron-star binaries:

\begin{enumerate}
    \item We compare the inferred $f_2(m)$ relation to the contact frequency in binary neutron star mergers to assess whether resonance with the orbital frequency is achieved~\cite{HoLai1999,BernuzziNagar2014};
    \item We estimate $f_2$ for the neutron-star components of binary neutron star and neutron star--black hole mergers observed by the LIGO-Virgo-KAGRA network~\cite{LIGOScientificCollaborationAasi2015,AcerneseAgathos2015,AkutsuAndo2021};
    \item We assess the distinguishability of $f$-mode dynamical tides in a realistic population of binary neutron star mergers detectable with existing~\cite{ObservingScenarios} and planned~\cite{MaggioreVanDenBroeck2020,EvansCorsi2023} gravitational-wave observatories.
\end{enumerate}

Our results are useful for evaluating prospects for observing dynamical tidal effects in neutron star binary inspirals, and for informing detector design for dedicated high-frequency sensitivity to postmerger oscillations~\cite{MartynovMiao2019,AckleyAdya2020,SrivastavaDavis2022}. Compared to existing studies of the distinguishability of $f$-mode dynamical tides~\cite{WilliamsPratten2022,PrattenSchmidt2020}, we account for EOS uncertainty via our marginalization procedure rather than fixing the EOS. We provide tabulated $f_2(m)$ sequences for $3 \times 10^4$ samples from the data-driven EOS posterior~\footnote{\url{doi:10.5281/zenodo.13952437}}, which can be used with existing $\Lambda(m)$ sequences from the same distribution to marginalize over quadrupolar $f$-mode frequency uncertainty in waveform models, or indeed to translate a future $f_2$ measurement into constraints on the neutron star EOS~\cite{PrattenSchmidt2022,PradhanGhosh2024} and composition~\cite{PradhanChatterjee2021,ShirkePradhan2024}.

The paper is structured as follows: Sec.~\ref{sec:fmode} describes the observationally conditioned ensemble of EOSs used in this work and sketches out the calculation of $f$-mode frequencies; Sec.~\ref{sec:results} presents the results of these calculations; and Sec.~\ref{sec:discussion} interprets and discusses the results. Details about the method of solution for the $f$-modes are provided in Appendix~\ref{sec:method}. In Appendix~\ref{sec:cowling}, we investigate the validity of the short-wavelength (Cowling) approximation~\cite{cowling1941non,finn1988relativistic} for the calculation of $f$-mode frequencies in the region of EOS parameter space preferred by the observations.

\section{Methodology} \label{sec:fmode}

Fundamental-mode oscillations are calculated as a dynamical perturbation of a static, spherically symmetric neutron star solution determined by the Tolman–Oppenheimer–Volkoff (TOV) equations of relativistic stellar structure~\cite{Tolman1939,OppenheimerVolkoff1939}, supplemented by an EOS for the neutron-star matter. The spacetime metric and fluid perturbations are governed by the Einstein field equations and the relativistic Euler equations. Here we describe our choices for the EOS and our method of solution for the perturbed Einstein-Euler system.

\subsection{Neutron star EOS}

We adopt a phenomenological, data-driven model for the unknown neutron star EOS. We use the nonparametric model of Ref.~\cite{LegredChatziioannou2021}, originally developed in Refs.~\cite{LandryEssick2019,EssickLandry2020}, which represents the supranuclear EOS as a Gaussian process conditioned on multi-messenger neutron star observations. These observations include the masses of the heaviest known pulsars~\cite{AntoniadisFreire2013,FonsecaCromartie2021}, the tidal deformability measurements from GW170817~\cite{GW170817} and GW190425~\cite{GW190425}, and the X-ray pulse profile modeling of PSR J0030+0451 and PSR J0740+6620~\cite{MillerLamb2019,MillerLamb2021}. We label these observational datasets PSR, GW and NICER, respectively. The Gaussian process is conditioned on the observations using the Bayesian inference framework developed in Ref.~\cite{LandryEssick2020}.

Sets of $10^4$ EOS samples~\footnote{\url{https://zenodo.org/records/6502467}} drawn from the posterior distributions informed by PSR, PSR+GW, and PSR+GW+NICER data, respectively, in Ref.~\cite{LegredChatziioannou2021} serve as Monte Carlo samples in our inference. By calculating the quadrupolar $f$-mode frequencies for a stable sequence of neutron stars for each EOS, we map the posterior distribution over the EOSs into one over $f_2(m)$. Evaluating the distribution for a specific value of $m$ produces a posterior on $f_m := f_2(m)|_m$ at the given mass scale. Generalizing to a mass measurement $P(m|d)$ for a given neutron star, the posterior on the star's quadrupolar $f$-mode frequency is $P(f_2|d) = \int dm P(f_2|m) P(m|d)$.

\subsection{Fundamental modes}

To derive the system of equations governing the $f$-mode oscillations of a nonrotating neutron star, we follow Ref.~\cite{thorne1967non} and consider a dynamical polar (even-parity) perturbation of a static, spherically symmetric spacetime with harmonic time dependence. The line element for the perturbed spacetime reads

\begin{align}
&ds^{2} = \nonumber \\ &-e^{\nu}\left(1+r^{\ell} H_{0} e^{i \omega t} Y_{\ell m}\right) d t^{2}
+e^{\lambda}\left(1-r^{\ell} H_{0} e^{i \omega t} Y_{\ell m}\right) d r^{2} \nonumber \\
&+\left(1-r^{\ell} K e^{i \omega t} Y_{\ell m}\right) r^{2} d \Omega^{2}
-2 i \omega r^{\ell+1} H_{1} e^{i \omega t} Y_{\ell m} d t d r
\end{align}
in the Regge-Wheeler gauge. The functions $\lambda(r)$, $\nu(r)$ are determined by the TOV equations; the perturbation is described by radial functions $H_0(r)$, $H_1(r)$, $K(r)$ and a complex oscillation mode frequency $\omega$. The real part of $\omega$ represents the angular frequency of the (quasi-)normal mode oscillation and the imaginary part is the mode damping factor. The angular dependence of the perturbation is encoded in spherical harmonics $Y_{\ell m}(\theta, \phi)$.

The accompanying fluid perturbations are described by a Lagrangian displacement vector whose spatial components are decomposed as 

\begin{subequations}
\begin{align}
   \xi^{r} = &  r^{l-1}e^{\lambda/2} W e^{i \omega t} Y_{\ell m} , \\
   \xi^{\theta} = & -r^{l-2} V e^{i \omega t} \partial_{\theta} Y_{\ell m} , \\
   \xi^{\phi} = & -\frac{r^{l-2}}{\sin^2{\theta}}V e^{i \omega t} \partial_{\phi}Y_{\ell m} ,
\end{align}
\end{subequations}
and a Lagrangian pressure perturbation

\begin{equation}
    \Delta p = -r^l e^{-\nu/2} X e^{i \omega t} Y_{\ell m} ,
\end{equation}
with radial functions $V(r)$, $W(r)$, $X(r)$. The corresponding Lagrangian density perturbation $\Delta \varepsilon = (d\varepsilon/dp) \Delta p$ is determined by the EOS. The metric perturbations are continuous across the stellar surface $R := r(p=0)$, while the fluid perturbations vanish outside the star.

Inserting the metric and fluid variables into the Einstein-Euler equations and linearizing in the perturbation produces a system of four coupled first-order ordinary differential equations for $H_1$, $K$, $W$, $X$, plus algebraic equations for $H_0$ and $V$. These equations, along with the TOV equations, are numerically integrated from $r=0$ to $r=R$. In vacuum, outside the star, the equations simplify significantly and can be reduced to a single second-order ordinary differential equation---the Zerilli equation---for the Zerilli-Moncrief function $Z$, which is related to the metric functions $H_1$ and $K$. The surficial values of the metric functions determine the initial condition for $Z$, so that the Zerilli equation can be numerically integrated from $r=R$ to infinity, or some appropriately large distance cutoff $r_\infty \gg R$.

To solve for the quadrupolar $f$-mode oscillations, following Refs.~\cite{lindblom1983quadrupole, 1985}, we specialize the perturbations to $\ell = 2$ and make a Fourier decomposition of $Z$ into ingoing and outgoing modes, turning the Zerilli equation into an eigenvalue problem for the complex mode frequency $\omega$. We search for the purely outgoing solution with zero radial nodes, i.e.~the eigenvalue with the smallest real part (lowest frequency). The quadrupolar $f$-mode frequency is calculated from the complex eigenfrequency as $f_2 = \mathrm{Re}(\omega)/2\pi$.

\section{Results} \label{sec:results}

\subsection{$f$-mode frequency vs mass relation} \label{subsec:contact}

\begin{figure}
    \centering
    \includegraphics[width=\linewidth]{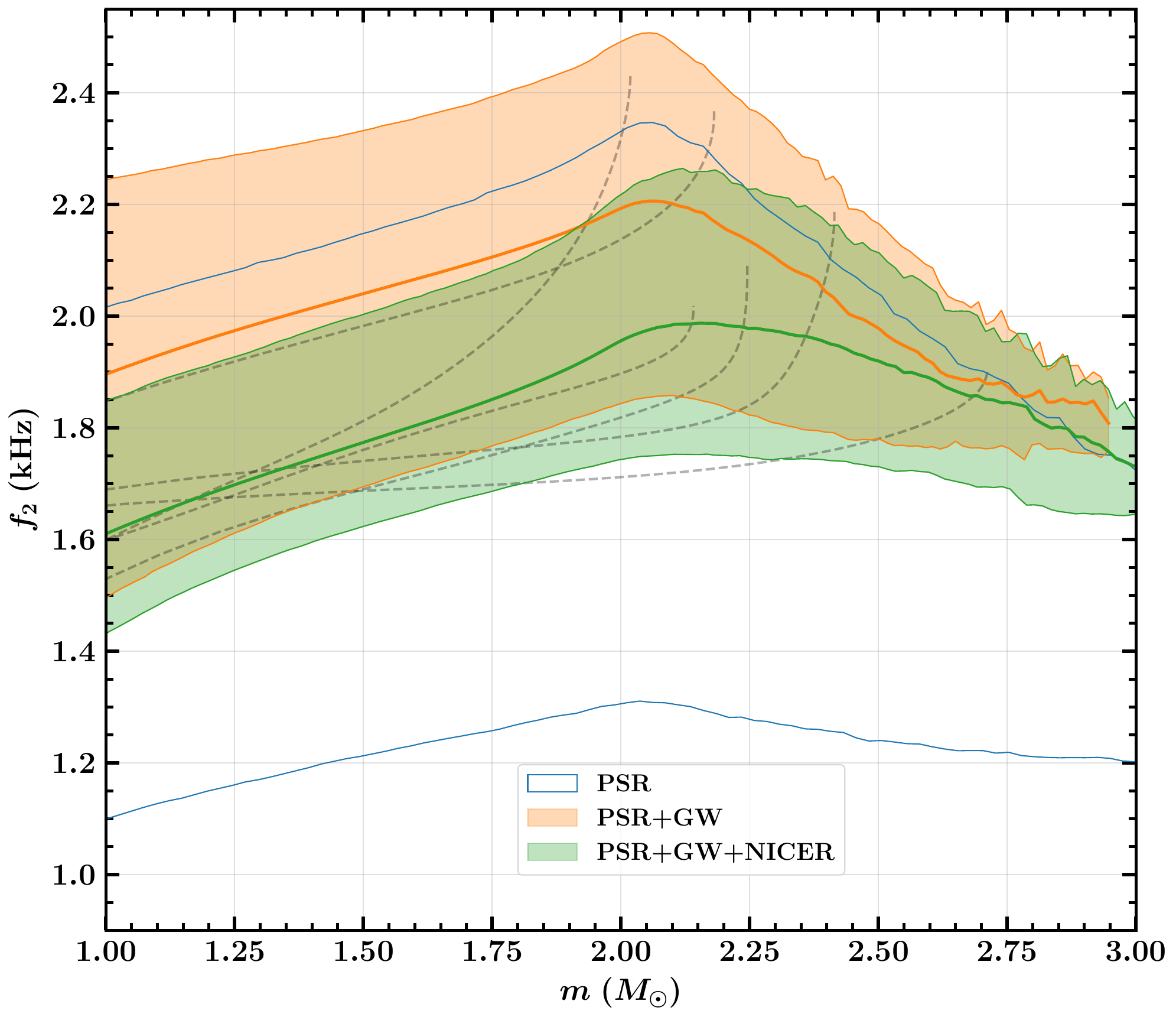}
    \caption{Mean (solid) quadrupolar $f$-mode frequency and symmetric 90\% credible interval (shaded) as a function of neutron star mass, as inferred from different astrophysical datasets. For the PSR dataset, we show only the 90\% credible interval. Every realization (dashed) of the $f_2(m)$ relation is monotonically increasing, but the mean of the distribution turns over above 2 $M_\odot$ because the EOSs that support larger neutron star masses are stiffer than average and thus have systematically lower $f_2$.
    }
    \label{fig:Envelope Plot}
\end{figure}

\begin{figure}
    \centering
    \includegraphics[width=\linewidth]{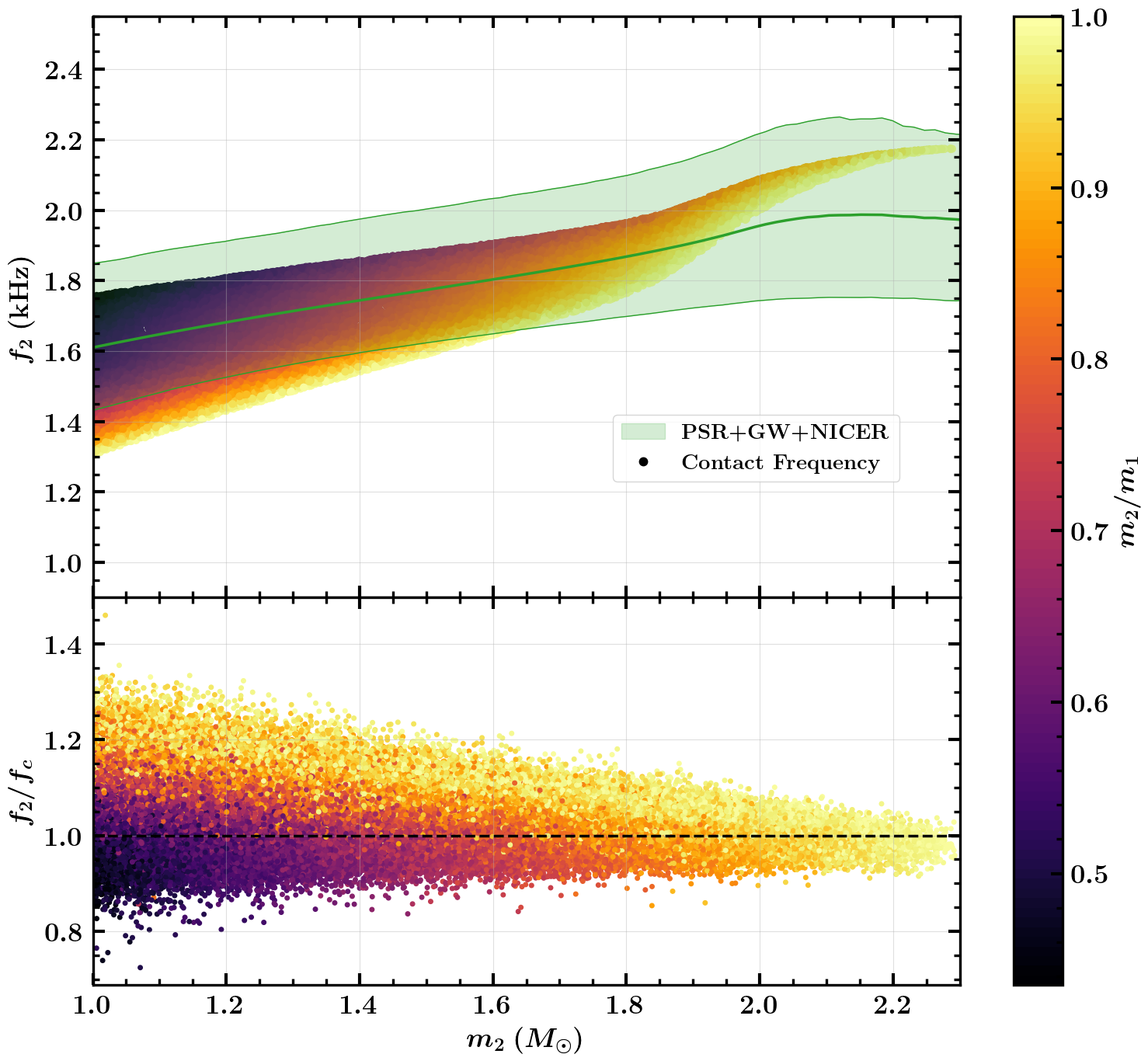}
    \caption{Comparison between the inferred quadrupolar $f$-mode frequency and the gravitational-wave frequencies at contact for a realistic population of binary neutron star coalescences. The PSR+GW+NICER constraints from Fig.~\ref{fig:Envelope Plot} are shown. Resonance of the orbit with the $f$-mode only occurs if $f_2$ is below the contact frequency.
    }
    \label{fig:contact}
\end{figure}

Computing the quadrupolar $f$-mode frequency as a function of neutron star mass for each EOS realization in the PSR, PSR+GW and PSR+GW+NICER datasets, we infer the distribution of $f_2(m)$ relations illustrated in Fig.~\ref{fig:Envelope Plot}, which depicts the mean and 90\% symmetric credible interval. The relation for each EOS is truncated at its maximum TOV mass, $ M_{\rm TOV} $. The constraints are progressively refined as the different observations---PSR, GW and NICER---are incorporated. The broad extent of the PSR-conditioned distribution reflects the large uncertainty in the EOS when predicated only on the existence of $2\,M_\odot$ pulsars, besides basic physical considerations like causality and thermodynamic stability. The addition of GW data shifts and tightens the distribution significantly, excluding stiffer EOSs that favor lower $f$-mode frequencies, especially at the intermediate mass scale probed by GW170817. These stiff EOSs are unable match the gravitational-wave measurement of neutron star tidal deformability~\cite{GW170817_EOS}. The inclusion of NICER data further shifts and narrows the distribution, particularly impacting the upper bound on $f_2(m)$: the stellar radii inferred from the pulse profile modeling exclude softer EOSs that produce more compact neutron stars with higher $f$-mode frequencies.

Within each dataset, we observe that the mean of the $f_2(m)$ distribution increases up to $\sim 2\,M_\odot$, the minimum TOV mass compatible with the PSR data, and decreases thereafter. This is a feature of the distribution over $f_2(m)$ only, and not of individual $f_2(m)$ relations: as the $f$-mode frequency is intrinsically linked to the stellar compactness---which influences the speed at which perturbations propagate, and always increases with neutron star mass---$f_2(m)$ is a monotonically increasing function. The apparent decline in the mean $f_2(m)$ beyond $2\,M_\odot$ is simply a result of the diminishing number of EOSs capable of supporting such a large $M_{\rm TOV}$ being systematically stiffer than average.

Using our EOS-informed estimate of the $f_2(m)$ relation, we investigate whether the orbit can achieve resonance with the $f$-mode during binary inspiral. Such resonances enhance the tidal phasing of the gravitational waveform, leaving a potentially measurable signature of dynamical tides~\cite{HindererTaracchini2016,SteinhoffHinderer2016,SchmidtHinderer2019,SteinhoffHinderer2021}. The dominant (22,2) resonance occurs when the orbital frequency is twice the quadrupolar $f$-mode frequency of one of the neutron stars~\citep{Lai1994,HoLai1999}, i.e.~when the GW frequency is equal to $f_2^{(1)}$ or $f_2^{(2)}$, where the superscripts indicate the binary component. This resonance condition is satisfied if $f_2^{(1,2)}$ is less than the GW frequency at merger, which we approximate as the Keplerian GW contact frequency~\citep{Agathos_2015} 

\begin{equation} \label{contact}
f_c = \sqrt{\frac{G (m_1 + m_2)}{\pi^2 (R_1 + R_2)^3}}
\end{equation}
for NS radii $R_{1,2}$. 

Fig.~\ref{fig:contact} shows our comparison between the inferred $f_2(m)$ relation and the distribution of contact frequencies for a realistic population of binary neutron star mergers. Our simulated population is characterized by a uniform neutron star mass distribution ranging from 1 $M_\odot$ to 2.3 $M_\odot$, the mean $M_{\rm TOV}$ for the PSR+GW+NICER EOS set, with random pairing into binaries. This population model is consistent with the neutron star mass distribution inferred from gravitational-wave observations to date~\cite{LandryRead2021,O3bPop}. Across the neutron star mass spectrum, the $f$-mode frequency is seen to be comparable to the contact frequencies attained in the population. Nonetheless, the resonance condition may be satisfied for the most unequal-mass mergers, where the secondary-component $f$-mode frequency is the lowest for a given total mass. For example, this is the case for a 2.0 $M_\odot$--1.2 $M_\odot$ configuration. This finding suggests that it is possible that resonance is achieved prior to merger for about 25\% of the merging binary neutron star population, even in the absence of rotational corrections to the $f$-mode frequency~\cite{Lai1994,HoLai1999}. 

\subsection{$f$-mode frequency estimates for individual neutron stars}

\begin{figure}
    \centering
    \includegraphics[width=\linewidth]{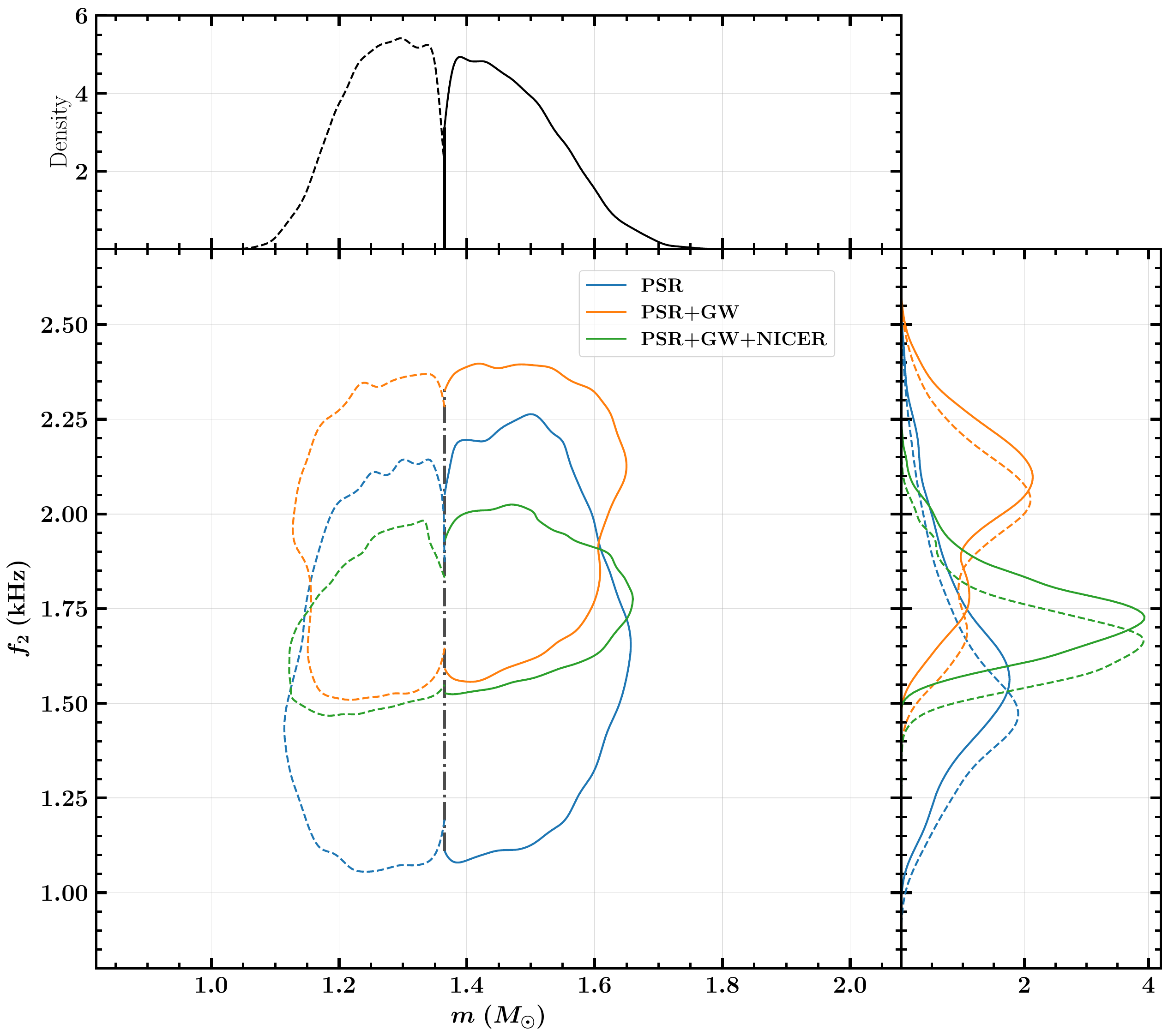}%
    \caption{Inferred two-dimensional marginal posterior distributions over mass and quadrupolar $f$-mode frequency for the primary (solid) and secondary (dashed) components of GW170817. The distributions are conditioned on three different observational datasets. Contours delimit 90\% credibility regions.
    }
    \label{fig:GW170817_GW190425_2D_posterior}
\end{figure}

\begin{figure}
    \centering
    \includegraphics[width=\linewidth]{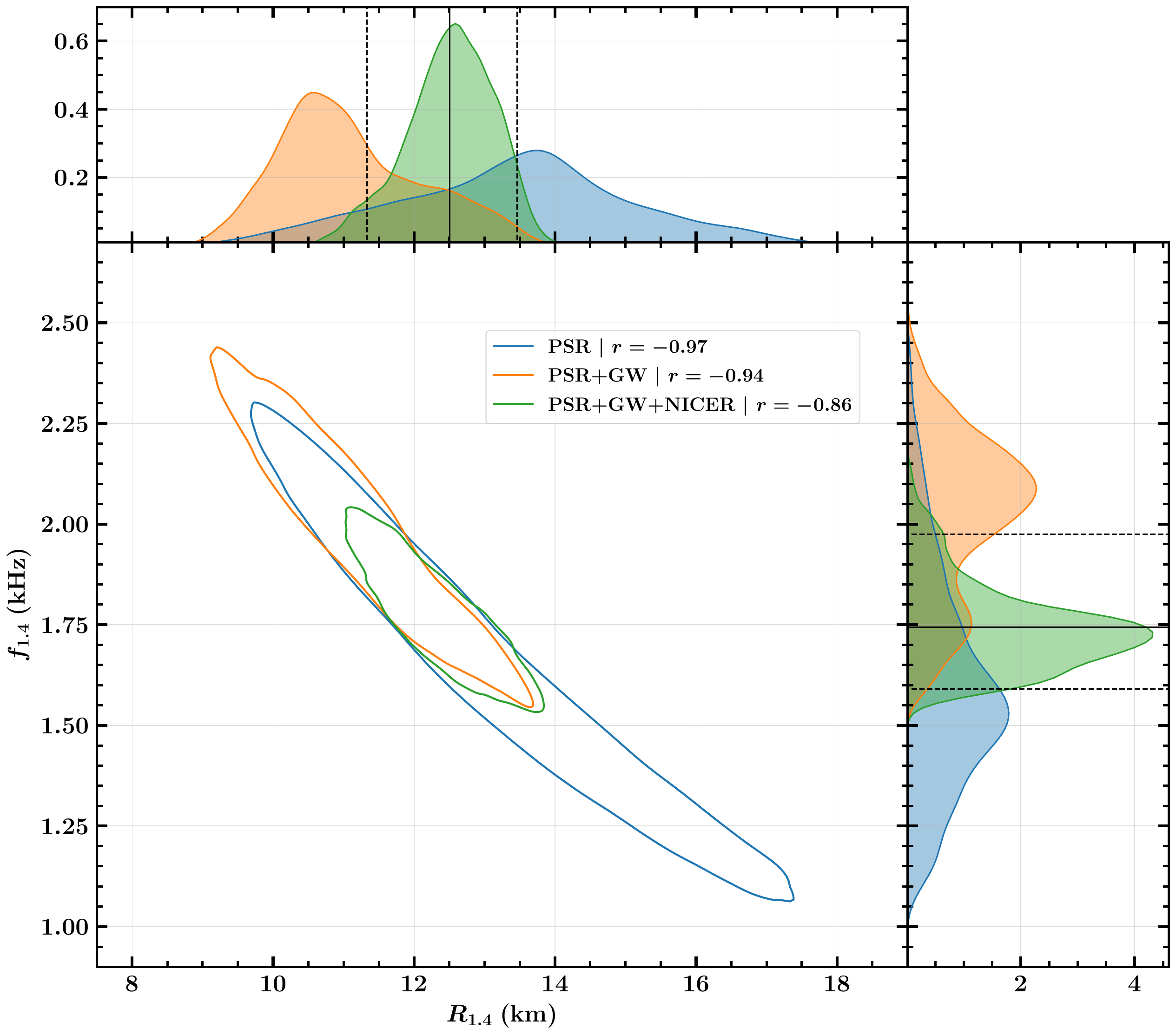} \\
    \includegraphics[width=\linewidth]{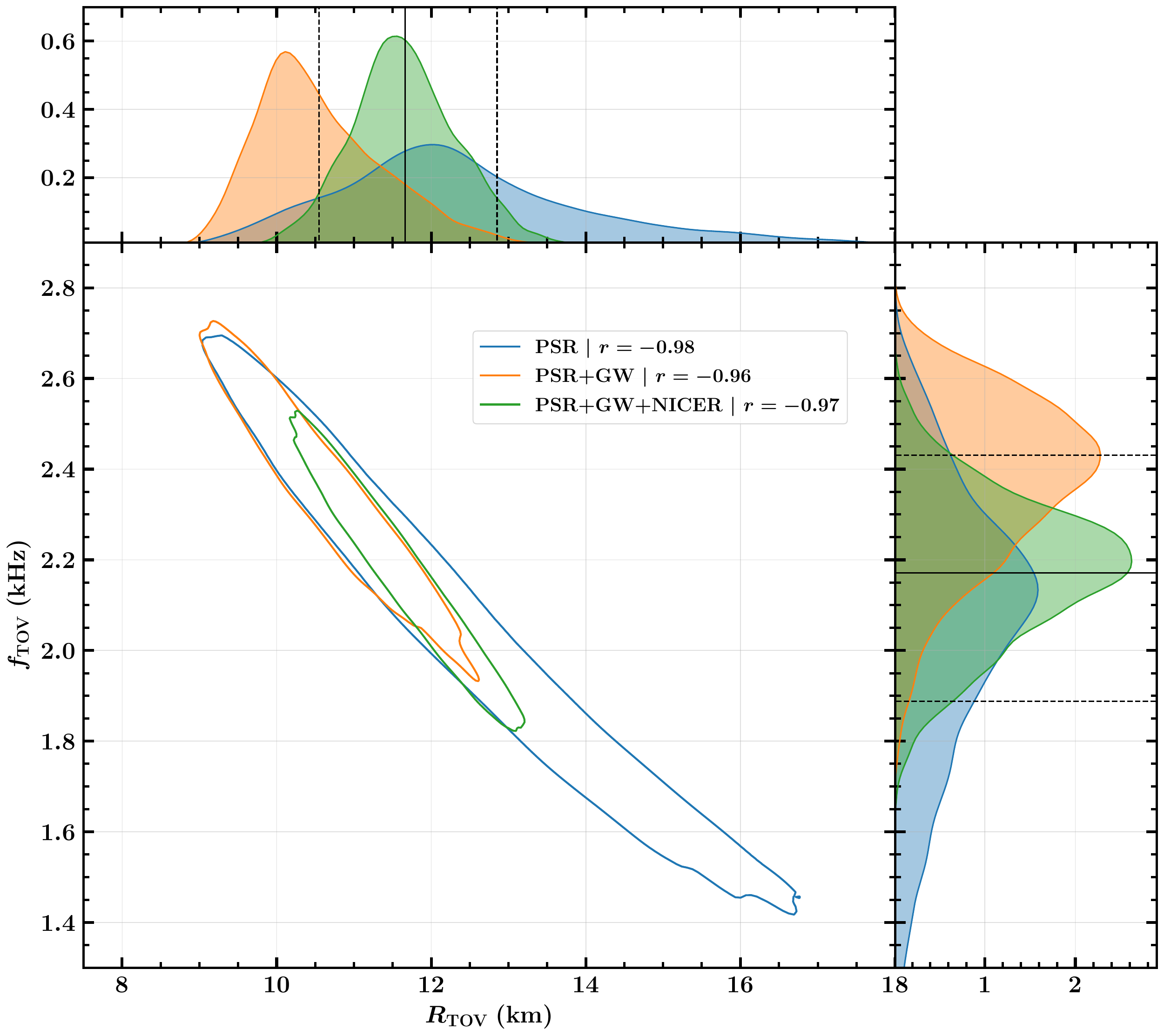}
    \caption{Inferred two-dimensional marginal posterior distributions over the radius of canonical (top) or TOV-mass (bottom) neutron stars and the corresponding quadrupolar $f$-mode frequency. The distributions are conditioned on three different observational datasets. Contours delimit 90\% credibility regions. The value of $f_{1.4}$ is strongly (anti-)correlated with $R_{1.4}$. A similar correlation is observed between $f_{\rm TOV}$ and $R_{\rm TOV}$. The correlation coefficients are quoted in the legend.
    }
    \label{fig:nsfmodes}
\end{figure}

Evaluating the posterior on $f_2(m)$ at a given value of $m$, or marginalizing the distribution over an uncertain mass measurement, we estimate the quadrupolar $f$-mode frequency for several specific neutron stars, as informed by each observational dataset. In Tables~\ref{tab:nsfmodes2} and~\ref{tab:nsfmodes}, we list the estimated $f_2$ value for canonical $1.4 M_\odot$ neutron stars, TOV stars, and the neutron star components of the compact binary mergers observed by the LIGO-Virgo-KAGRA collaboration to date. In particular, we consider the primary and secondary components of the binary neutron star mergers GW170187~\cite{GW170817_Discovery,GW170817} and GW190425~\cite{GW190425}, as well as the secondary components of the neutron-star--black-hole mergers GW200105, GW200115 and GW230529 \cite{GW200105_GW200115,GW230529}. Quoted symmetric 90\% credible intervals about the mean account for the uncertainty in the $f_2(m)$ relation. The error in $f_{\rm TOV}$ also incorporates the uncertainty in the value of $M_{\rm TOV}$ inferred from each dataset.

Figure~\ref {fig:GW170817_GW190425_2D_posterior} focuses on the predictions for GW170817, illustrating the two-dimensional marginalized posterior distributions over the masses $ m_{1,2} $ and quadrupolar $ f $-mode frequencies $ f_2^{(1,2)} $ of the compact binary's components. The posteriors are conditioned on the three different observational datasets, progressively incorporating PSR, GW, and NICER observations. Each observational dataset incrementally improves the precision of the $f_2$ estimates. The hierarchy $m_1 \geq m_2$ of the component masses enforces a hierarchy $f_2^{(1)} \geq f_2^{(2)}$, which is not apparent in the contour plot or the one-dimensional marginal distributions, when a common EOS is imposed.

We also investigate the correlations between pairs of these observables. The marginal two-dimensional posteriors on $R_{1.4}$ and $f_{1.4}$, and on $R_{\rm TOV}$ and $f_{\rm TOV}$, are plotted in Fig.~\ref{fig:nsfmodes} for the three observational datasets considered. We observe a strong (anti-)correlation between $R_{1.4}$ and $f_{1.4}$ in the top panel: larger-radius neutron stars are less compact, giving them lower $f$-mode frequencies. The constraining effect of the GW (respectively, NICER) observations, relative to the PSR data, is to disfavor EOSs that prefer physically larger (smaller) neutron stars. Because of the strong correlation with $f_{1.4}$, these observations significantly limit the range of allowed values for the quadrupolar $f$-mode frequency of a canonical neutron star. The lower panel exhibits a similarly strong (anti-)correlation between $ R_{\text{TOV}} $ and $ f_{\text{TOV}}$.

\begin{table}
    \centering
    \caption{Inferred mean and 90\% credible interval for the quadrupolar $f$-mode frequency of a canonical 1.4 $M_\odot$ neutron star and a maximum-mass TOV star, as informed by different observational datasets. Constraints on the canonical radius $R_{1.4}$ and the TOV mass $M_{\rm TOV}$ are also given.
    }
    \setlength{\tabcolsep}{2mm}
    \renewcommand{\arraystretch}{2}
    \scalebox{0.8}{
        \begin{tabular}{lccc}
            \hline \hline 
            Dataset & PSR & PSR+GW & PSR+GW+NICER \\
            \hline 
            $R_{1.4}$ (km) & $13.44^{+2.92}_{-2.94}$ & $11.11^{+1.92}_{-1.48}$ & $12.51^{+0.95}_{-1.18}$ \\
            $f_{1.4}$ (kHz) & $1.60^{+0.52}_{-0.41}$ & $2.01^{+0.31}_{-0.35}$ & $1.74^{+0.23}_{-0.15}$ \\
            $M_{\rm TOV}$ ($M_\odot$) & $2.33^{+0.59}_{-0.30}$ & $2.21^{+0.27}_{-0.18}$ & $2.25^{+0.35}_{-0.22}$ \\
            $R_{\rm TOV}$ (km) & $12.47^{+3.48}_{-2.48}$ & $10.58^{+1.59}_{-1.16}$ & $11.66^{+1.19}_{-1.11}$ \\
            $f_{\rm TOV}$ (kHz) & $2.06^{+0.46}_{-0.55}$ & $2.37^{+0.27}_{-0.34}$ & $2.17^{+0.26}_{-0.28}$ \\
            \hline \hline
        \end{tabular}}
    \label{tab:nsfmodes2}
\end{table}

\begin{table*}
    \centering
    \caption{Estimated quadrupolar $f$-mode frequency for neutron stars in LIGO-Virgo-KAGRA compact binary mergers, as informed by different observational datasets. The mean and 90\% credible interval for $f_2$ are quoted for the primary and secondary components of the binary neutron star mergers GW170817 and GW190425, as well as the secondary components of the neutron-star--black-hole mergers GW200105, GW200115 and GW230529.
    }
    \setlength{\tabcolsep}{3mm}
    \renewcommand{\arraystretch}{2}
    \scalebox{0.8}{
        \begin{tabular}{lcccccccc}
            \hline \hline 
            & $m_{1}$ ($M_\odot$) & $m_{2}$ ($M_\odot$) & \multicolumn{2}{c}{PSR} & \multicolumn{2}{c}{PSR+GW} & \multicolumn{2}{c}{PSR+GW+NICER} \\
            \hline 
            & & & $f_{2}^{(1)}$ (kHz) & $f_{2}^{(2)}$ (kHz) & $f_{2}^{(1)}$ (kHz) & $f_{2}^{(2)}$ (kHz) & $f_{2}^{(1)}$ (kHz) & $f_{2}^{(2)}$ (kHz) \\
            \hline 
            GW170817 & $1.48^{+0.14}_{-0.11}$ & $1.26^{+0.09}_{-0.11}$ & $1.63^{+0.52}_{-0.41}$ & $1.56^{+0.52}_{-0.39}$ & $2.02^{+0.31}_{-0.36}$ & $1.95^{+0.34}_{-0.38}$ & $1.75^{+0.23}_{-0.16}$ & $1.68^{+0.24}_{-0.16}$ \\
            GW190425 & $2.12^{+0.46}_{-0.42}$ & $1.33^{+0.27}_{-0.23}$ & $1.84^{+0.50}_{-0.50}$ & $1.58^{+0.54}_{-0.42}$ & $2.16^{+0.32}_{-0.35}$ & $1.97^{+0.34}_{-0.38}$ & $1.96^{+0.29}_{-0.24}$ & $1.70^{+0.24}_{-0.19}$ \\
            GW200105 & -- & $1.91^{+0.34}_{-0.24}$ & -- & $1.83^{+0.54}_{-0.48}$ & -- & $2.18^{+0.32}_{-0.37}$ & -- & $1.93^{+0.29}_{-0.23}$ \\
            GW200115 & -- & $1.54^{+0.75}_{-0.40}$ & -- & $1.66^{+0.56}_{-0.45}$ & -- & $2.04^{+0.36}_{-0.40}$ & -- & $1.78^{+0.34}_{-0.23}$ \\
            GW230529 & -- & $1.48^{+0.51}_{-0.29}$ & -- & $1.64^{+0.55}_{-0.44}$ & -- & $2.03^{+0.35}_{-0.39}$ & -- & $1.76^{+0.29}_{-0.20}$ \\
            \hline \hline
        \end{tabular}}
    \label{tab:nsfmodes}
\end{table*}

\subsection{Distinguishability of $f$-modes}

Informed by our best current knowledge of the neutron star EOS and mass distribution, we investigate the detectability of $f$-mode dynamical tides in a simulated population of binary neutron star coalescences. The tidal excitation of the $f$-mode oscillation by a binary companion's gravitational influence leads to additional phasing of the inspiral gravitational waveform, relative to the case where only adiabatic tides are active. The tidal phase contributed by quadrupolar $f$-mode dynamical tides is given analytically as

\begin{align} \label{psidyn}
    \Psi_{\rm dyn} = &-\frac{1}{96}(10\pi\sqrt{3} - 27 - 30\log{2}) (155-147 X_1)  \nonumber \\
    &\times \frac{\Lambda_1 {X_1}^5}{X_2{\Omega_1}^2} \left(\frac{v}{c}\right)^{11} + (1 \leftrightarrow 2)
\end{align}
in Ref.~\cite{SchmidtHinderer2019}'s \textsc{fmtidal} frequency-domain model. Here, $X_{1,2} = m_{1,2}/M$ express the mass ratio relative to the total mass $M = m_1+m_2$, $\Omega_{1,2} = 2\pi G m_{1,2} f_{1,2}/c^3$ are the dimensionless angular $f$-mode frequencies, and $(v/c)^2 = (\pi GM f_{\rm gw}/c^3)^{2/3}$ is the post-Newtonian (PN) parameter.

\begin{figure}
    \centering
    \includegraphics[width=1.1\linewidth]{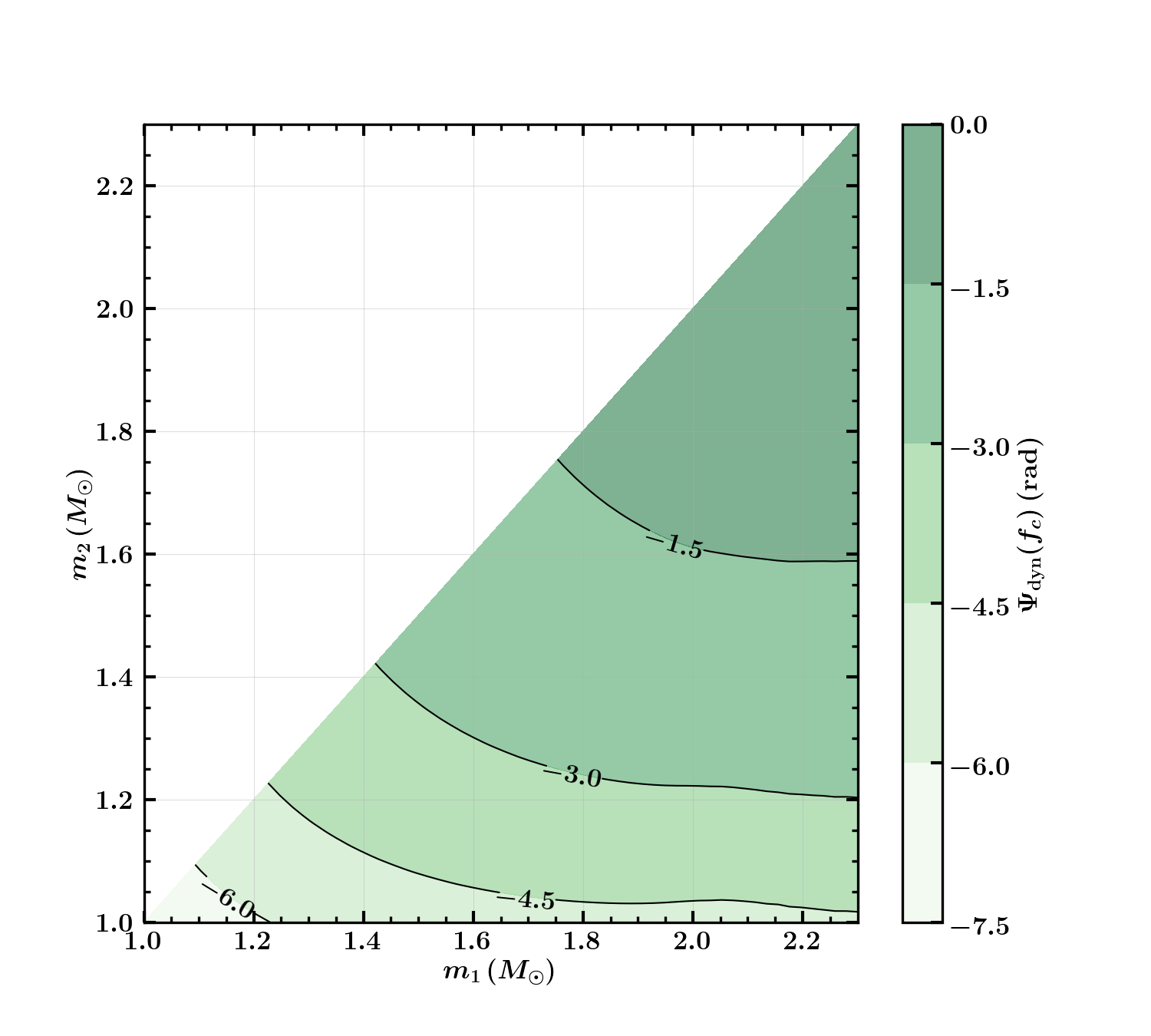}%
    \caption{EOS-averaged tidal phase at merger from quadrupolar $f$-mode dynamical tides as a function of binary neutron star component masses. The tidal phase is evaluated at the contact frequency with the \textsc{fmtidal} model from Ref.~\cite{SchmidtHinderer2019}. The EOS distribution is informed by the PSR+GW+NICER dataset.
    }
    \label{fig:phase}
\end{figure}

We evaluate Eq.~\eqref{psidyn} at contact for every binary neutron star in the simulated population described in Sec.~\ref{subsec:contact}. As the \textsc{fmtidal} model for dynamical tides ignores spin effects, we take the neutron star to be nonspinning. We average over the PSR+GW+NICER EOS set when calculating the tidal deformabilities $\Lambda_{1,2}$ (which enter into $\Psi_{\rm dyn}$ directly) and the neutron star radii $R_{1,2}$ (which enter into $f_c$). The contours of constant $\Psi_{\rm dyn}(f_c)$ in the resulting distribution are plotted as a function of component masses in Fig.~\ref{fig:phase}. For our assumed population model and EOS posterior, the mean dynamical tidal phase at contact is $|\Psi_{\rm dyn}(f_c)| = 2.8^{+2.8}_{-2.0}$ rad. The largest tidal phase occurs for the binaries with the lowest total mass in our population, namely 1 $M_\odot$--1 $M_\odot$ coalescences, for which $|\Psi_{\rm dyn}(f_c)| = 7.4^{+2.4}_{-2.6}$ rad, where the error bars account for EOS uncertainty.

Given these predictions for the inspiral tidal phasing, we quantify the ability of different gravitational-wave detectors to distinguish the effect of quadrupolar $f$-mode dynamical tides on top of the adiabatic tidal deformations captured by $\Lambda$, following Ref.~\cite{WilliamsPratten2022}. We consider four detector configurations: Advanced LIGO operating at design sensitivity, Advanced LIGO operating at enhanced A+ sensitivity~\cite{ObservingScenarios}, Einstein Telescope in its triangular 10-km configuration~\cite{MaggioreVanDenBroeck2020}, and Cosmic Explorer in its 40-km configuration~\cite{EvansCorsi2023}.

For every binary neutron star coalescence in the simulated population, we compute the distinguishability signal-to-noise ratio (SNR)

\begin{equation}
\rho_{\rm dyn} = \sqrt{D/2(1-\mu_{\rm dyn})}
\end{equation}
for the $f$-mode dynamical tides in terms of the match

\begin{equation}
    \mu_{\rm dyn} = \frac{\langle h|h_{\rm dyn} \rangle}{\sqrt{\langle h|h\rangle \langle h_{\rm dyn}|h_{\rm dyn} \rangle}} 
\end{equation}
between an inspiral waveform with ($h_{\rm dyn}$) and without ($h$) $f$-mode dynamical tides; $D$ is the number of intrinsic parameters in the waveform, including $f_2^{(1,2)}$. The notation $\langle \; | \; \rangle$ represents a noise-weighted inner product

\begin{equation}
    \langle h_1|h_2\rangle = 4 \, \mathrm{Re} \int^{f_c}_{f_{\rm min}} df \, h_1 h_2^*/S_n
\end{equation}
of frequency-domain waveforms $h_1(f)$ and $h_2(f)$ relative to the power spectral density $S_n(f)$ of the gravitational-wave detector, with ${}^*$ denoting complex conjugation. The integral is computed from a minimum gravitational-wave frequency, which we take to be $f_{\rm min} = 10$ Hz, to the gravitational-wave contact frequency $f_c$, which approximates the frequency at merger. We use the \textsc{TaylorF2\_Tidal} waveform model~\cite{BuonannoIyer2009,WadeCreighton2014} in our implementation, which includes adiabatic quadrupolar tidal effects captured by $\Lambda_{1,2}$ at 5PN. To incorporate the effect of $f$-mode dynamical tides, we augment the phase of the waveform according to Eq.~\eqref{psidyn}. We consider the inference of $D = 6$ intrinsic parameters, namely the component masses, tidal deformabilities and the $f$-mode frequencies $f_2^{(1,2)}$.

\begin{figure}
    \centering
    \includegraphics[width=1.1\linewidth]{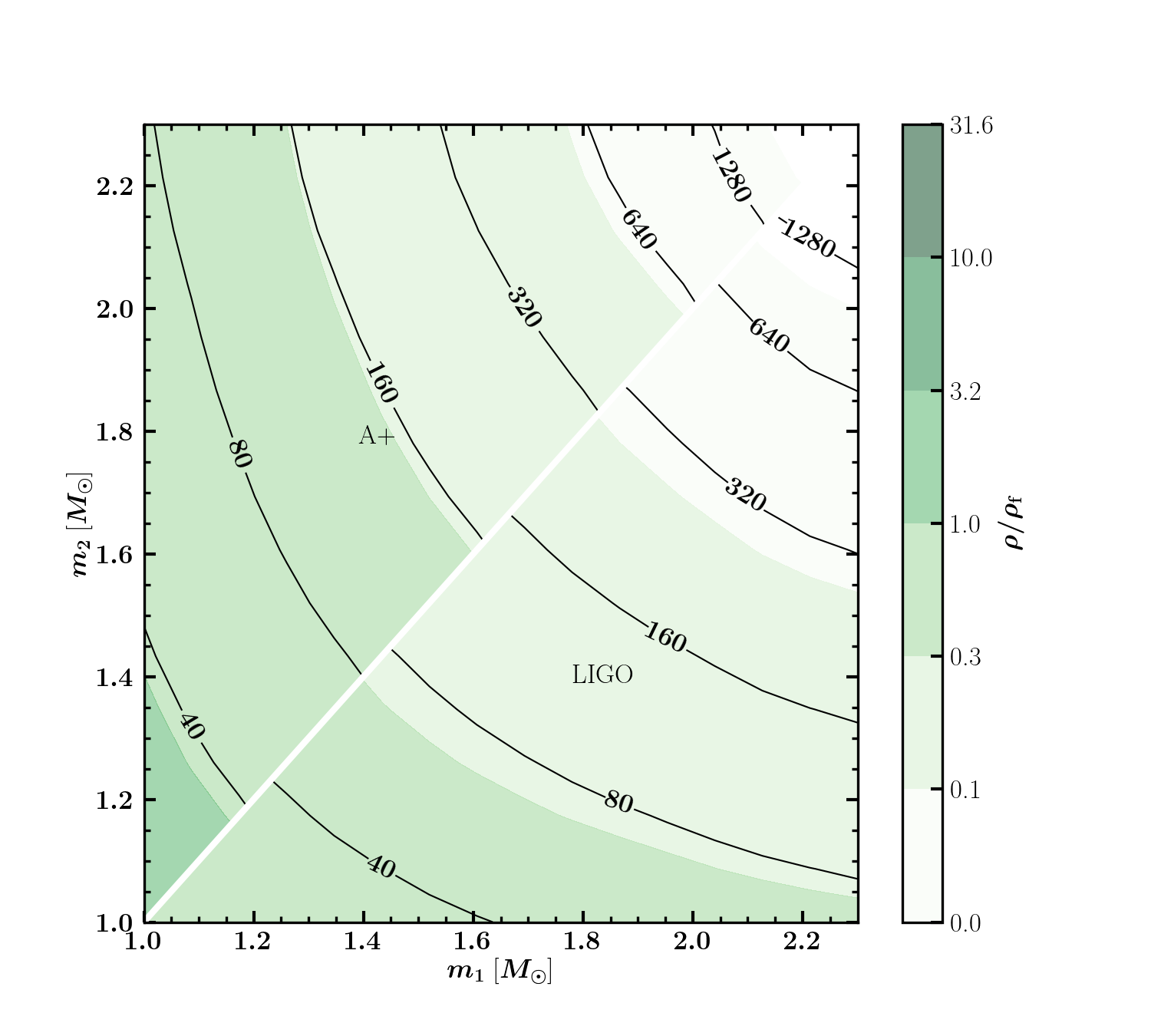} \\
    \includegraphics[width=1.1\linewidth]{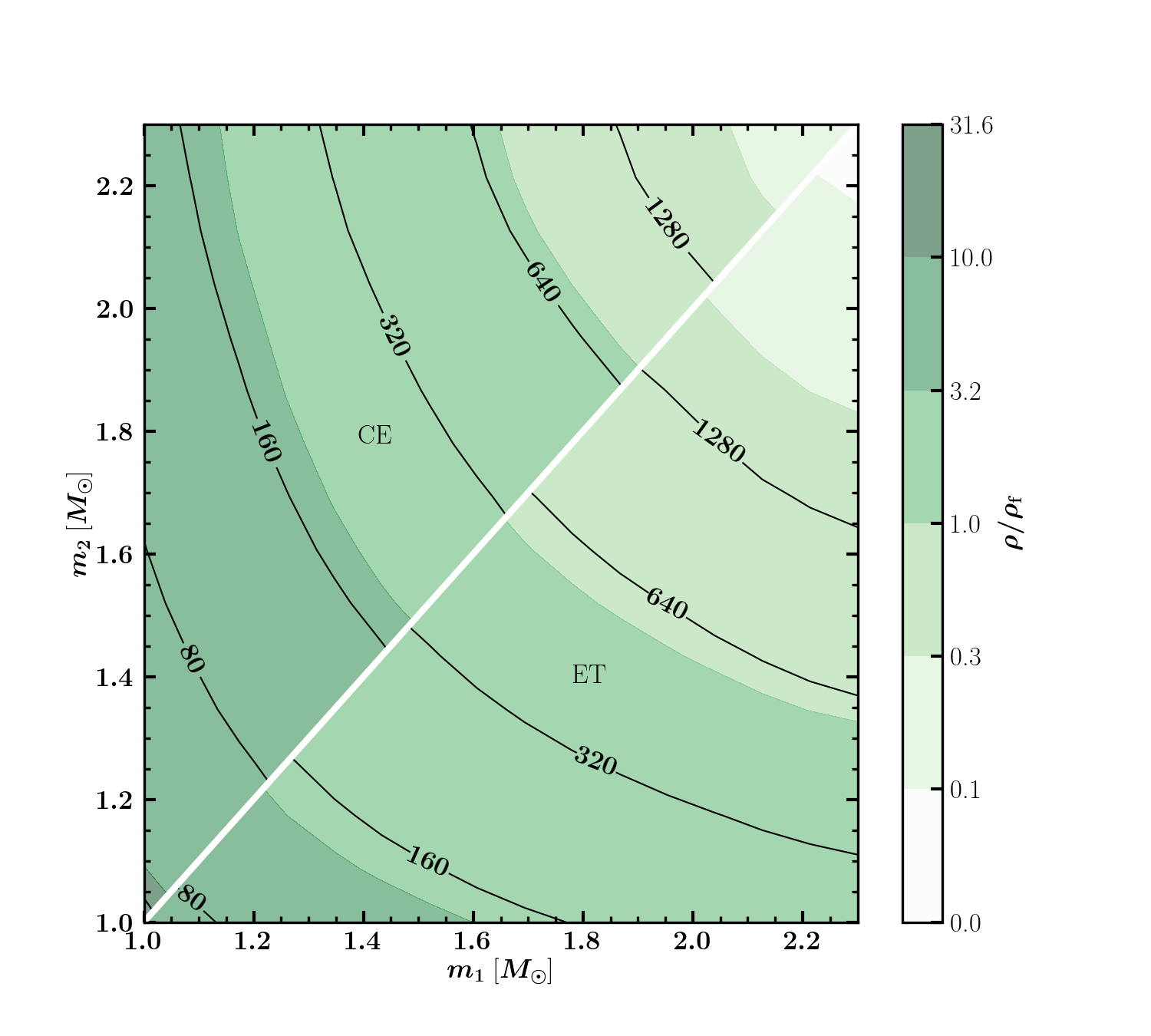}
    \caption{Signal-to-noise ratio required to distinguish $f$-mode dynamical tides from adiabatic tides in a binary neutron star inspiral, averaged over EOS uncertainty, as a function of component masses $m_{1,2}$. The solid contours show the distinguishability SNR $\rho_{\rm dyn}$, while the shaded contours show the ratio $\rho/\rho_{\rm dyn}$ with the optimal SNR for an optimally oriented source at 100 Mpc. The lower (upper) triangular region of the top panel shows the results for LIGO at design (A+) sensitivity, while the lower (upper) triangular region of the bottom panel shows the same for Einstein Telescope (Cosmic Explorer). The EOS distribution is informed by the PSR+GW+NICER dataset.
    }
    \label{fig:distinguishability}
\end{figure}

The impact of the $f$-mode oscillations on the inspiral phase is distinguishable when the optimal SNR $\rho = \sqrt{\langle h | h \rangle}$ for the coalescence surpasses the distinguishability SNR, i.e.~when $\rho/\rho_{\rm dyn} > 1$. We plot this ratio, averaged over PSR+GW+NICER EOS uncertainty, as a function of binary neutron star component masses for optimally oriented sources at 100 Mpc, as seen by the various detectors, in Fig.~\ref{fig:distinguishability}. We remark that none of the coalescences satisfy the distinguishability criterion at LIGO design sensitivity; at A+ sensitivity, only those with total mass $M \lesssim 2.4\,M_\odot$ do at this distance. The situation is more promising with a next-generation detector: dynamical tides are distinguishable for more than half of the optimally oriented sources at 100 Mpc in Einstein Telescope (Cosmic Explorer). In particular, the coalescences with $M \lesssim 3.4\,M_\odot$ ($3.8\,M_\odot$) all have $\rho/\rho_{\rm dyn} > 1$ at this distance.

As an alternative, distance-independent metric, we also plot contours of constant $\rho_{\rm dyn}$, averaged over PSR+GW+NICER EOS uncertainty, as a function of the binary component masses. We see that an optimal SNR of 75 (90) is required to distinguish $f$-mode dynamical tides in a typical 1.4 $M_\odot$--1.4 $M_\odot$ coalescence with LIGO at design (A+) sensitivity. With a next-generation detector, an optimal SNR of 270 (155) in Einstein Telescope (Cosmic Explorer) suffices to distinguish them in the same coalescence. 
As binary neutron star mergers with network SNRs of several hundreds will routinely be detected in the era of next-generation gravitational-wave observatories~\cite{EvansAdhikari2021,BorhanianSathyaprakash2022,GuptaAfle2023}, this suggests that the observation of dynamical tidal effects will become routine in the future, although they are largely inaccessible to current-generation detectors like Advanced LIGO and Virgo.

\section{Discussion} \label{sec:discussion}

From our results above, we conclude that our best EOS-informed estimate of the $f_2(m)$ relation is low enough in frequency that resonance between the $f$-mode and the orbit can be achieved in $\sim 25\%$ of binary neutron star inspirals---even without significant neutron star spins, which reduce the $f$-mode frequency via rotational corrections~\cite{Lai1994,HoLai1999}. We estimate the size of $\Psi_{\rm dyn}(f_c)$, the tidal phase accumulated from dynamical tides, to be as large as $7^{+2}_{-3}$ rad at merger in these cases. This effect is not distinguishable from adiabatic tidal phasing for a binary neutron coalescence observed at a typical distance of 100 Mpc with design-sensitivity Advanced LIGO, but may become marginally distinguishable at A+ sensitivity, and will be comfortably so with next-generation detectors like Einstein Telescope and Cosmic Explorer.

Our distinguishability study of the $f$-mode dynamical tides is inspired by Ref.~\cite{WilliamsPratten2022}'s, and arrives at qualitatively similar findings in the scenarios that are common to both works. In particular, our EOS-averaged results are most comparable to the case where the soft APR4 EOS is assumed. However, unlike Ref.~\cite{WilliamsPratten2022}, we only consider the quadrupole dynamical tides; we neglect the octupole contribution since it is suppressed by two PN orders. Moreover, we study the distinguishability for Advanced LIGO, and when considering Cosmic Explorer, we use an updated (somewhat less optimistic) sensitivity curve~\cite{EvansCorsi2023}. Our results are most directly comparable in the context of the Einstein Telescope, where we find distinguishability SNRs about 1.5 times higher than for APR4. 

The findings from our comparison between the $f_2(m)$ relation and the distribution of contact frequencies in the binary neutron star merger population are sensitive to our assumptions about the EOS. In particular, our fiducial results are based on the PSR+GW+NICER EOS set, which includes X-ray pulse profile modeling constraints on the radii of PSR J0030+0451 and PSR J0740+6620. However, several competing analyses of the NICER data for these pulsars exist~\cite{RileyWatts2019,VinciguerraSalmi2024,RileyWatts2021,SalmiVinciguerra2022,DittmannMiller2024,SalmiChoudhury2024}, some of which find different preferred hotspot geometries, and consequently different radius solutions. This raises a question about the robustness of the NICER radius estimates. If we were to conservatively use the PSR+GW-conditioned $f_2(m)$ relation in place of the PSR+GW+NICER-conditioned one in our comparison with the contact frequencies, we would conclude that the quadrupolar $f$-mode frequencies are too high for resonance with the orbit to be achieved before merger. On the other hand, our treatment of the $f$-mode oscillations in this paper neglects the effects of neutron star spin, which reduce the mode frequency, thereby increasing the chance that the mode is resonantly excited before merger.

Our constraints on the $f_2(m)$ relation can be used to predict the quadrupolar $f$-mode frequency for individual neutron stars. This may be relevant for gravitational-wave detectors being designed for dedicated high-frequency sensitivity to postmerger gravitational waves~\cite{MartynovMiao2019,AckleyAdya2020,SrivastavaDavis2022}; peaks in the postmerger spectrum are often contextualized in terms of a corrected fundamental mode oscillation.

Although measuring the $f$-mode frequencies directly will likely have to wait until next-generation ground-based gravitational-wave detectors arrive on the scene, neglecting $f$-mode dynamical tides may bias EOS inference even with a current-generation population of binary neutron star merger observations~\cite{PrattenSchmidt2022}. Current waveform models, such as the \textsc{SEOBNRTv4} surrogate~\cite{LackeyPurrer2019} and \textsc{IMRPhenomPv2\_NRTidalv3}~\cite{AbacDietrich2024}, avoid this by incorporating $f$-mode dynamical tides, while fixing the unknown $f$-mode frequencies via an empirical relation that connects them to the adiabatic tidal deformability $\Lambda$. By making our EOS-informed set of $f_2(m)$ relations available, we hope to enable an alternative prescription: direct marginalization of the $f$-mode frequencies over an EOS distribution conditioned on the desired suite of astronomical observations. This would have the advantage of incorporating uncertainty in the $f$-mode frequency vs tidal deformability correlation into the waveform models, rather than simply fixing the frequencies to a point-estimate.

\acknowledgments
P.L. is supported by the Natural Sciences \& Engineering Research Council of Canada (NSERC). NSERC also supports U.M. through a Discovery Grant (RGPIN-2023-03346). Research at Perimeter Institute is supported in part by the Government of Canada through the Department of Innovation, Science and Economic Development and by the Province of Ontario through the Ministry of Colleges and Universities. B.K. acknowledges partial support from the Department of Science and Technology, Government of India, with grant no. CRG/2021/000101. This material is based upon work supported by NSF's LIGO Laboratory which is a major facility fully funded by the National Science Foundation. The authors are grateful for computational resources provided by the LIGO Lab and supported by NSF Grants PHY-0757058 and PHY-0823459.

\appendix
\setcounter{equation}{0}
\renewcommand{\theequation}{A\arabic{equation}}
\section{Computation of $f$-mode frequencies} \label{sec:method}

Our strategy for computing the quadrupolar $f$-mode frequencies is as follows. The Einstein-Euler system for the metric and fluid perturbations defined in Sec.~\ref{sec:fmode} results in four first-order coupled ordinary differential equations,

\begin{widetext}
    \begin{align} 
        \frac{d H_{1}}{d r} =& -\left[l+1+2 b e^{\lambda}+ 4 \pi r^{2} e^{\lambda}(p-\varepsilon)\right] \frac{H_{1}}{r} +\frac{e^{\lambda}}{r}\left[H_{0}+K-16 \pi(\varepsilon+p) V\right] , \label{odei} \\[8pt]
        \frac{d K}{d r} =& \frac{H_{0}}{r}+\left(n_{l}+1\right) \frac{H_{1}}{r} +\left[e^{\lambda} \mathrm{Q}-l-1\right] \frac{K}{r}-8 \pi(\varepsilon+p) e^{\lambda / 2} \frac{W}{r} , \\[8pt]
        \frac{d W}{d r} =& -\frac{(l+1)}{r}\left[W+l e^{\lambda/2} V\right] +re^{\lambda / 2}\left[\frac{e^{-\nu / 2} X}{(\varepsilon+p)c_{ad}^2}+\frac{H_{0}}{2}+K\right] , \\[8pt]
        \frac{d X}{d r} =& \frac{-l}{r} X +\frac{(\varepsilon+p) e^{\nu / 2}}{2r} \left\{\left(1-e^{\lambda} \mathrm{Q}\right) H_{0}+\left(r^{2} \omega^{2} e^{-\nu}+n_{l}+1\right) H_{1}\right.+\left(3 e^{\lambda} \mathrm{Q}-1\right) K \nonumber \\[2pt]
        & -\frac{4\left(n_{l}+1\right) e^{\lambda} \mathrm{Q}}{r^{2}} V-2\left[\omega^{2} e^{\lambda / 2-\nu}\left. + 4 \pi(\varepsilon+p) e^{\lambda / 2}-r^{2} \frac{d}{d r}\left(\frac{e^{\lambda / 2} \mathrm{Q}}{r^{3}}\right)\right] W\right\} \label{odef}
    \end{align}  
\end{widetext}
and two algebraic equations,

\begin{align} 
H_{0}  = & \left\{ \left[n_{l}-\omega^{2} r^{2} e^{-\nu}-\mathrm{Q}\left(e^{\lambda} \mathrm{Q}-1\right)\right]K\right.  \nonumber \\
& -\left[\left(n_{l}+1\right) \mathrm{Q}-\omega^{2} r^{2} e^{-(\nu+\lambda)}\right] H_{1} \nonumber \\
&\left.+ 8 \pi  r^{2} e^{-\nu / 2} X   \right\}\left(2 b+n_{l}+\mathrm{Q}\right)^{-1} , \label{ae1} 
\end{align}
\begin{align} 
 V &= \left[\frac{X}{\varepsilon+p}-\frac{Q}{r^{2}} e^{(\nu+\lambda) / 2} W-e^{\nu / 2} \frac{H_{0}}{2}\right] \frac{e^{\nu / 2}}{\omega^{2}} \label{ae2}
\end{align}
which govern the perturbations inside the star.  $H_0$, $H_1$, $K$ describe metric perturbations, whereas $W$, $X$, $V$ pertain to fluid perturbations; $n_l= \frac{1}{2}\left(l-1\right)\left(l+2\right)$, $Q = b + 4 \pi r^2 p$, and $b = Gm/c^2r$. The metric functions $\lambda(r)$, $\nu(r)$ and fluid variables $m(r),p(r),\epsilon(r)$ are determined by the background solution of the TOV equations, and are subsequently employed to numerically integrate the aforementioned differential equations. When solving the differential Eqs. (\ref{odei})–(\ref{odef}) in conjunction with the algebraic Eqs. (\ref{ae1}) and (\ref{ae2}), it is imperative to impose appropriate boundary conditions, such that the perturbation functions remain finite throughout the interior of the star (particularly at the center \( r = 0 \)), and the perturbed pressure \(\Delta p\) vanishes at the surface. Function values at the center of the star can be determined utilizing the Taylor series expansion method, as described in Appendix B of Ref. \cite{lindblom1983quadrupole}:

\begin{equation}
\begin{aligned}
X(0) =& \left\{\left[\frac{4 \pi}{3} \left(\varepsilon_{0}+3 p_{0}\right)-\frac{\omega^{2}}{l} e^{-\nu_{0}}\right] W(0)\right. \\ & \left.+\frac{K(0)}{2}\right\}\left(\varepsilon_{0}+p_{0}\right) e^{\nu_{0} / 2} ,
\end{aligned}
\end{equation}
\begin{equation}
H_{1}(0)=\frac{l K(0)+8 \pi\left(\varepsilon_{0}+p_{0}\right) W(0)}{n_{l}+1} ,
\end{equation}
\begin{equation}
H_0(0) =  K(0) , \quad W(0)=  1 ,
\end{equation}
where $p_0= p(r=0)$, $\varepsilon_0= \varepsilon(r=0)$ and $\nu_0= \nu(r=0)$. The vanishing perturbed pressure at the stellar surface is equivalent to the condition \( X(R) = 0 \), since \(\Delta p = -r^l e^{-\nu/2} X\). This boundary condition is achieved by solving for two trial solutions with $K(0)=\pm (\epsilon_o + p_o)$ and then linearly constructing the correct solution satisfying  \( X(R) = 0 \).

In the region outside the star, the perturbations are described by the Zerilli equation~\cite{1971}. Setting $m=M$ (since $m(r>R)=M$), $W=V=X=0$ (as no fluid is present outside the star), and replacing $H_1,K$ with new variables $Z$, $dZ/dr^*$ defined by

\begin{equation} \label{zerilli_ic}
\left(\begin{array}{c}
K(r) \\
H_{1}(r)
\end{array}\right)=\left(\begin{array}{lc}
g(r) & 1 \\
h(r) & k(r)
\end{array}\right)\left(\begin{array}{c}
Z\left(r^{*}\right) / r \\
d Z\left(r^{*}\right) / d r^{*} ,
\end{array}\right)
\end{equation}
where

\begin{equation}
\begin{aligned}
g(r) &=\frac{n_l(n_l+1)+3 n_l b+6 b^{2}}{(n_l+3 b)} , \\
h(r) &=\frac{\left(n_l-3 n_l b-3 b^{2}\right)}{(1-2 b)(n_l+3 b)} , \\
k(r) &=\frac{1}{1-2 b} ,
\end{aligned}
\end{equation}
we recover the second-order differential equation

\begin{equation} \label{ze}
\frac{d^{2} Z}{d r^{* 2}}=\left(V_{Z}(r^*)-\omega^{2}\right) Z(r^*)
\end{equation}
for the Zerilli-Moncrief function $Z$. Here, the effective potential $V_{Z}$ is defined as

\begin{equation}
V_{Z}(r^*)=(1-2 b) \frac{2 n_l^{2}(n_l+1)+6 n_l^{2} b+18 n_l b^{2}+18 b^{3}}{{r^*}^{2}(n_l+3 b)^{2}} ,
\end{equation}
with $b = M/r$ and $r^* = r + 2M\log(\frac{r}{2M}-1)$. The Zerilli-Moncrief function $Z$ describes the metric perturbations outside the star, including gravitational radiation in the far field. $Z$ can be decomposed into incoming ($Z_+$) and outgoing ($Z_-$) modes, related by complex conjugation, as~\cite{1971}

\begin{equation} \label{dze}
\binom{Z(\omega)}{d Z / d r^*}= \left(\begin{array}{cc}
Z_{-}(\omega) & Z_{+}(\omega) \\
d Z_{-} / d r^* & d Z_{+} / d r^*
\end{array}\right)\binom{A_{-}(\omega)}{A_{+}(\omega)} . \\
\end{equation}
For $r \gg R$, the amplitudes $A_{+}(\omega)$ and $A_{-}(\omega)$ of the incoming and outgoing modes can be expanded in powers of $r$ as

\begin{align}
Z_- = & e^{-i\omega r^*} \left[ a_0 + \frac{a_1}{r} + \frac{a_2}{r^2} + ...\right], \nonumber \\
\frac{dZ_-}{dr^*} = & -i\omega e^{-i\omega r^*} \left[ a_0 + \frac{a_1}{r} + \frac{a_2 + i\alpha_1(1 - 2b)/\omega}{r^2} + ... \right];
\end{align}
the expansion coefficients satisfy the recursion relation~\cite{Chandrasekhar}

\begin{align}
\alpha_1 =& -i(n_l+1)\alpha_0, \nonumber \\
\alpha_2 =& \frac{-[n_l(n_l+1) + i\text{M}(3/2+3/n_l)]\alpha_0}{2\omega^2} ,
\end{align}
with $\alpha_0$ an arbitrary complex number representing an overall phase. The mode solutions must satisfy the correct boundary condition at $r=\infty$: the amplitude of incoming radiation $A_{+}(\omega)$ vanishes for physical eigenmodes, such that the gravitational radiation is purely outgoing in the far field. 

To determine $\omega$, we initially perform a numerical integration of the Zerilli equation from $r=R$ to a specified distance cutoff $r_\infty \gg R$, which we designate as $r_\infty = 25\omega^{-1}$. Subsequently, we align the solution of Eq.~\eqref{ze} with the expansion Eq.~\eqref{dze} in the far field to obtain the mode amplitudes $A_{+}$ and $A_{-}$. We proceed to identify the root of $A_{+}(\omega)=0$ in the complex plane \cite{lindblom1983quadrupole}. Given that the imaginary component of the eigenfrequency is typically small (less than $10^{-3}$ times the magnitude of the real component) for the $f$-mode, we approximate the complex eigenfrequency by modeling $A_{+}$ near the eigenfrequency as a quadratic

\begin{equation}
    A_{+}(\omega) \approx A_0+A_1 \omega+A_2 \omega^2 ,
\end{equation}
where $A_0$, $A_1$, and $A_2$ are complex constants determined by $A_{+}(\omega)$ along the real axis of $\omega$. The solution of this quadratic equation provides an estimate for both the real and imaginary parts of the eigenfrequency $\omega$. The $f$-mode frequency cited in the main text is $f_2 := \mathrm{Re}(\omega)/2\pi$. This process is reiterated for each EOS across a neutron star configuration mass grid comprising 40 points ranging from 1 $M_\odot$ to $M_{\rm TOV}$. Consequently, we generate a tabulated $f_2(m)$ relation for every EOS within the PSR, PSR+GW, and PSR+GW+NICER datasets. Approximately 3\% of the EOSs, characterized by significant density discontinuities leading to near-singular sound speed profiles, are excluded to mitigate substantial numerical errors in the $f$-mode frequency computation, which is highly susceptible to variations in sound speed.

\renewcommand{\theequation}{B\arabic{equation}}
\section{Fundamental modes in the Cowling approximation} \label{sec:cowling}

\begin{figure}
    \centering
    \includegraphics[width=\linewidth]{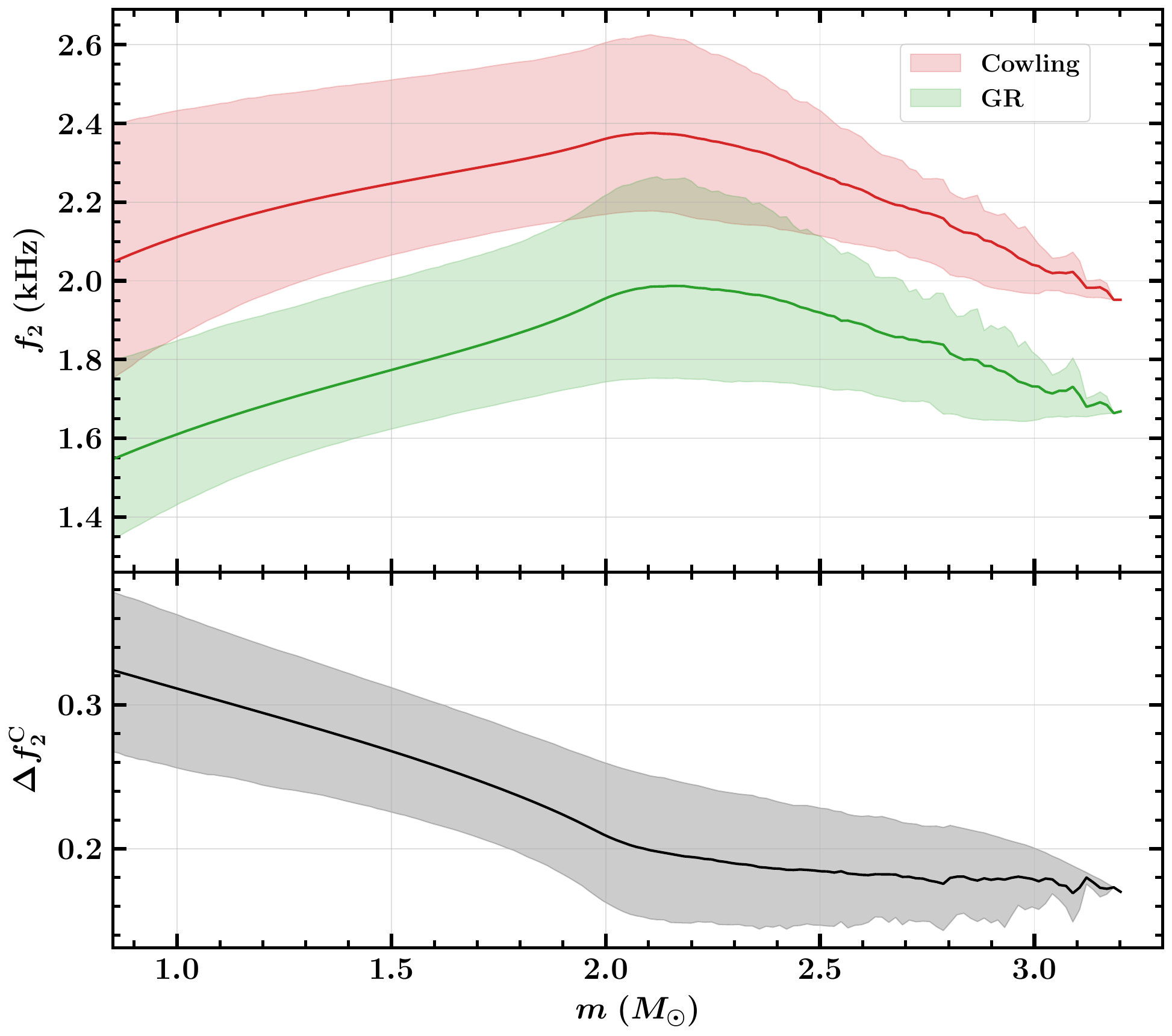}%
    \caption{Comparison between quadrupolar $f$-mode frequencies computed in perturbative general relativity and in the Cowling approximation. The top panel compares the distribution of $f_2(m)$ relations for the PSR+GW+NICER dataset from Fig.~\ref{fig:Envelope Plot} (green) against the corresponding distribution computed in the Cowling approximation (red), with the shaded region encapsulating 90\% credible intervals on $f_2$. The bottom panel shows the fractional difference $\Delta f_2^{\rm C} := (f_2^{\rm C}/f_2 - 1)$ between the perturbative general-relativistic and Cowling results as a function of NS mass, with the shaded region again denoting 90\% credible intervals due to the EOS uncertainty in the PSR+GW+NICER dataset. The discrepancy is larger for lower NS masses.
    }
    \label{fig:cowling1}
\end{figure}

In the main part of this study, we conduct a perturbative general-relativistic analysis to compute the fundamental neutron-star oscillation mode. Owing to the significant computational cost associated with this method, a simplified alternative formalism, commonly employed in the literature to determine $f$-mode frequencies, is the Cowling approximation~\cite{cowling1941non, finn1988relativistic}. The Cowling approximation represents a short-wavelength approach that disregards the metric perturbations in the aforementioned Einstein-Euler framework. Consequently, this method involves solving fewer coupled equations and entails reduced computational runtime \cite{lindblom1990accuracy, doneva2012nonradial, samuelsson2007neutron}. It is well-documented that the oscillation mode frequencies derived using the Cowling approximation can deviate from those obtained through general-relativistic calculations by as much as 20\% to 30\%~\cite{yoshida1997accuracy, athulkp}, and tend to systematically overestimate the $f$-mode frequency \cite{Sotani_2020}. This section delineates the methodology underpinning the Cowling approximation and recomputes the quadrupolar $f$-mode frequencies within this framework to compare its predictions with the general-relativistic results presented in the main body of the paper.

Ignoring the metric perturbations in Eqs. (\ref{odei})–(\ref{odef}), i.e.~enforcing $H_1 = K = H_0 = 0$, we are left with simplified fluid perturbation equations~\cite{ZhaoLattimer2022}:

\begin{align} 
   r\frac{dW}{dr} = & -(l+1) \left[W - l e^{\nu + \lambda/2} U \right] \\ & - e^{\lambda/2}(\omega r)^2 \left[U - \frac{e^{\lambda/2}QW}{(\omega r)^2} \right] , \nonumber \\ r \frac{dU}{dr} = & e^{- \nu + \lambda/2 } \left[W - l e^{\nu - \lambda/2} U\right] ,
\end{align}
where we have defined $U = -e^{-\nu}V$. The initial condition for integration of the system starting at $r=0$ is~\cite{athulkp}

\begin{equation}
\left.\frac{W}{U}\right|_{r=0}=l e^{\nu(r=0)} ,
\end{equation}
where $\nu(r=0)$ is determined by the background TOV solution and $W(r=0)$ is arbitrary; the $f$-mode frequency depends only on the ratio $W/U$. For quadrupolar $f$-modes, we specialize to $\ell = 2$ and integrate numerically from $r=0$ to the stellar surface $r=R$. The oscillation mode frequency $\omega$ is computed by optimizing the following boundary conditions at the stellar surface~\cite{athulkp}:

\begin{equation}
\left.\frac{W}{U}\right|_{r=R}=\frac{\omega^{2} R^{3}}{M} \sqrt{1-\frac{2 M}{R}} .
\end{equation}
As in the main text, this procedure is repeated on a grid of neutron star configurations for each EOS in the PSR, PSR+GW and PSR+GW+NICER datasets to build up a posterior distribution over $f_2(m)$ relations.

We compare the distribution of quadrupolar $f$-mode frequency vs mass relations obtained within the Cowling approximation to that computed in perturbative general relativity in Fig.~\ref{fig:cowling1}. As shown in the top panel, the quadrupolar $f$-mode frequencies computed in the Cowling approximation are consistently higher across the neutron star mass spectrum when conditioned on the PSR+GW+NICER dataset. For instance, at a canonical neutron star mass of $ 1.4 \, M_\odot $, the Cowling approximation predicts $ f_{1.4} = 2.23^{+0.27}_{-0.19}$ kHz, whereas the general-relativistic calculation estimates $ f_{1.4} = 1.75^{+0.23}_{-0.15}$ kHz, a discrepancy of about 28\%. For $ 2.5 \, M_\odot $ neutron stars, the Cowling approximation estimates $f_2 = 2.27^{+0.16}_{-0.16}$ kHz, while the fully relativistic calculation yields $f_2 = 1.92^{+0.19}_{-0.19}$ kHz, with an 18\% deviation. The lower panel of Fig.~\ref{fig:cowling1} further illustrates this point by showing the fractional difference between the quadrupolar $f$-mode frequency in the Cowling approximation and the perturbative general-relativistic calculation, denoted as $\Delta f_2^{\rm C} := (f_2^{\rm C}/f_2 - 1)$. Somewhat counterintuitively, we find that the EOS-averaged fractional differences are most pronounced for lighter, and thus less compact, neutron stars. Nonetheless, the consistent 20--30\% discrepancy across the neutron star mass spectrum highlights the importance of the metric perturbations for accurate determination of $f$-mode frequencies.

\bibliographystyle{apsrev4-1}
\bibliography{nsfmode}
\end{document}